\documentclass{aastex63}

\usepackage{ulem}
\usepackage{mathrsfs}
\usepackage{amsmath}
\usepackage{graphicx}
\usepackage{subfigure}

\submitjournal{APJ}
\received{2019 December 1}
\revised{2020 February 14}
\accepted{2020 February 16}

\shorttitle{period analysis}
\shortauthors{Q-S., Wang; S-B., Qian; Z-T., Han; X-H., Fang; L., Zang and W., Liu}

\graphicspath{{./}{figures/}}

\begin{document}
\title{Spot Model for Identifications of Periods in Asynchronous Polars}

\correspondingauthor{Qishan, Wang}
\email{wangqs@ynao.ac.cn}

\author{Qishan, Wang}
\affiliation{Yunnan Observatories, Chinese Academy of Sciences (CAS), \\
P. O. Box 110, 650216 Kunming, China.}
\affiliation{Key Laboratory of the Structure and Evolution of Celestial Objects, \\
Chinese Academy of Sciences, P. O. Box 110, 650216 Kunming, China.}
\affiliation{University of Chinese Academy of Sciences, \\
Yuquan Road 19\#, Sijingshang Block, 100049 Beijing, China.}

\author{Shengbang, Qian}
\affiliation{Yunnan Observatories, Chinese Academy of Sciences (CAS), \\
P. O. Box 110, 650216 Kunming, China.}
\affiliation{Key Laboratory of the Structure and Evolution of Celestial Objects, \\
Chinese Academy of Sciences, P. O. Box 110, 650216 Kunming, China.}
\affiliation{University of Chinese Academy of Sciences, \\
Yuquan Road 19\#, Sijingshang Block, 100049 Beijing, China.}
\affiliation{Center for Astronomical Mega-Science, Chinese Academy of Science, \\
20A Datun Road, Chaoyang Distric, Beijing, 100012, P.R. China}

\author{Zhongtao, Han}
\affiliation{Yunnan Observatories, Chinese Academy of Sciences (CAS), \\
P. O. Box 110, 650216 Kunming, China.}
\affiliation{Key Laboratory of the Structure and Evolution of Celestial Objects, \\
Chinese Academy of Sciences, P. O. Box 110, 650216 Kunming, China.}
\affiliation{University of Chinese Academy of Sciences, \\
Yuquan Road 19\#, Sijingshang Block, 100049 Beijing, China.}

\author{Xiaohui, Fang}
\affiliation{Yunnan Observatories, Chinese Academy of Sciences (CAS), \\
P. O. Box 110, 650216 Kunming, China.}
\affiliation{Key Laboratory of the Structure and Evolution of Celestial Objects, \\
Chinese Academy of Sciences, P. O. Box 110, 650216 Kunming, China.}
\affiliation{University of Chinese Academy of Sciences, \\
Yuquan Road 19\#, Sijingshang Block, 100049 Beijing, China.}

\author{Lei, Zang}
\affiliation{Yunnan Observatories, Chinese Academy of Sciences (CAS), \\
P. O. Box 110, 650216 Kunming, China.}
\affiliation{Key Laboratory of the Structure and Evolution of Celestial Objects, \\
Chinese Academy of Sciences, P. O. Box 110, 650216 Kunming, China.}
\affiliation{University of Chinese Academy of Sciences, \\
Yuquan Road 19\#, Sijingshang Block, 100049 Beijing, China.}

\author{Wei, Liu}
\affiliation{Guizhou Provincial Key Laboratory of Radio Data Processing, School of Physics and Electronic Sciences, Guizhou Normal University, Guiyang 550001, people's Republic of China}

\begin{abstract}
We improved the discless accretion models of Wynn \& King, considering the effects of the changing aspect due to the white dwarf spin and the variable feeding intensity caused by the asynchronism, and set up a more general spot model which is not sensitive to the different forms of these effects and can be applied for the period analysis of the optical and X-ray light curve. The spot model can produce the power spectra compatible with the observations, and its simulations limit the ratio $P_{spin}/P_{orb}<2$ between the powers at the white dwarf spin and the binary orbital frequencies, which is a strong criterion for identification of periods. Then we recognize the periods for CD Ind, BY Cam and 1RXS J083842.1-282723. The spot model reveals a complex accretion geometry in the asynchronous polars, which may indicate that the complex magnetic field causes their asynchronism. We think 1RXS J083842.1-282723 is a pre-polars because of its highest asynchronism and stable light curve. Giving the unstable accretion process in asynchronous polars, the period analysis of the long-term light curve will make the orbital signal prominent. \end{abstract}


\section{Introduction} \label{sec:intro}
Polars are interacting semidetached binaries in which a red dwarf overflows its Roche lobe and transfers material to a highly magnetic ($B\gtrsim10MG$) white dwarf (WD). The matter will be captured at the threading point and, along the field lines, impact onto the WD surface, feeding the accretion regions near the magnetic poles. The regions are limited to a small area but contribute to most radiation \citep[see][]{Cropper1990Polars}. The field would force the WD to spin synchronously with the revolution of the system, whereas there are still four asynchronous polars (APs) well confirmed, including CD Ind, BY Cam, V1432 Aql and V1500 Cyg, with their angular velocities of the WDs more or less than their orbital ones. Although nova burst is blamed for the asynchronous rotation given the 1975 nova eruption of V1500 Cyg, there are no evidences of the existence of nova shells around other three ones \citep{Pagnotta2016Nonshells}. 

Due to the asynchronism, the WD will rotate slowly with respect to the red dwarf and so do the accretion regions, which cause a pole-switching modulation with each pole active for half a beat cycle \citep{Wynn1992power_spectra}. However, the models in \cite{Wynn1992power_spectra} were designed for the intermediate polars based on the X-ray radiation and did not include the variation of the feeding intensity which definitely occurs in the accretion processes of the APs (see Section \ref{sec:cdind} and Section \ref{sec:RXJ0838-2827}). In fact the coupling of the matter to WD field is complex and even unstable, and the light curves of these systems vary from one cycle to the next, especially that of BY Cam. These characteristics make the period identification confused owing to the lack of stable periodical signals, such as the eclipsing profiles in V1432 Aql and reflection effect in V1500 Cyg. While the accurate orbital period is very important not only for the study of the secular evolution of the cataclysmic variables but also for searching the extra-solar planets orbiting them by the eclipsing method \citep[e.g.][]{Qian2010DP_LEO, Qian2011HUAqr}.

In this work we improved the models in \cite{Wynn1992power_spectra} based on the observations and shown the predictions of this new models, which are described in Section \ref{sec:model}, then used the results to identify the periods of CD Ind (Section \ref{sec:cdind}), BY Cam (Section \ref{sec:bycam}) and 1RXS J083842.1-282723 (Section \ref{sec:RXJ0838-2827}), and finished with the discussion and summary in Section \ref{sec:Summary}. 

\section{Models}\label{sec:model}
The material overflows through the inner Lagrangian point (L1) and follows a free trajectory until it is captured by the magnetic filed lines at the threading point and impacts the WD atmosphere, producing a shock front near the magnetic poles. The shock region is the dominant source of optical and X-ray emission, and changes aspect due to the WD spin. Assuming a geometry in which the binary orbits in frequency $\Omega$ and the WD spins in frequency $\omega$ with its rotation axis perpendicular to the orbital plane, the radiation from a spot at co-latitude $m$, if visible, will produce cosine-like changes as 
\begin{equation}\label{eq:aspect_changing}
L\cos{\theta} = L(\cos{i}\cos{m}+\sin{i}\sin{m}\cos{\Delta\psi}),
\end{equation}
by the spherical cosine rule, here $L$ is the luminosity of the spot, $i$ is the inclination of the binary, and $\Delta\psi$ is the phase difference between the line of sight and the spot. Setting $\Delta\psi=0$ in the Eq.\ref{eq:aspect_changing} gives the maximum visible brightness $B=L\cos{(m-i)}$ for the spot, and this parameter is directly relative to the observation so is more practical.

Because of the asynchronism, the WD will slowly rotate with respect to the L1 and the matter will be fed toward to different regions dependent on the locations of the poles relative to the threading point. Also the distance of the accretion spot to the threading point will modulate the accretion rate - the feeding intensity - then the pole luminosity at the beat frequency of the WD spin frequency $\omega$ and the binary orbital frequency $\Omega$. This kind of modulation in the luminosity of a certain accretion region, denoted as Spot A, can be simply represented by 
\begin{equation}\label{eq:feeding_intensity}
\left(\frac{cos\Delta\psi_{AL}+1}{2}\right)^{\gamma},
\end{equation}
where $\Delta\psi_{AL}$ is the phase difference between the spot A and the L1, and $\gamma$ adjusts the scale, e.g. the luminosity of the pole will reduce to half if the phase difference is up to $\sim 33^{\circ}~(\sim 0.1$ in units of $2\pi)$ for $\gamma=8$. we set a lower limit $\gamma>3$ given that the luminosity is less than 0.1 of its maximum for the phase difference $\pi/2~(90^{\circ})$, and $\gamma=8$ was used in following simulations.

For the high-inclination systems, some parts of the accretion stream can pass in front of the accretion regions, absorb the flux and produce a broad trough in their light curve. These systems are more likely to be the eclipsing ones, such as V1432 Aql, of which the periods are easily concluded. And there are no evidences for the absorption effects contained in the systems discussed below. So the model do not include any effects of the absorption.

Without the loss of generality, each accretion region is viewed as point and their positions are not relevant. Then the geometry of APs is defined by the inclination $i$, the WD spin frequency $\omega$, the binary orbital frequency $\Omega$, the initial phase $\phi_L$ for the L1, and initial phase $\phi$, co-latitude $m$ and luminosity $L$ for each accretion spot. Here $\phi$ is the initial longitude of the spot with respect to the projection of the line of sight on the orbital plane, and the direction of WD spin is considered as positive.

\begin{figure}[ht!]
\plotone{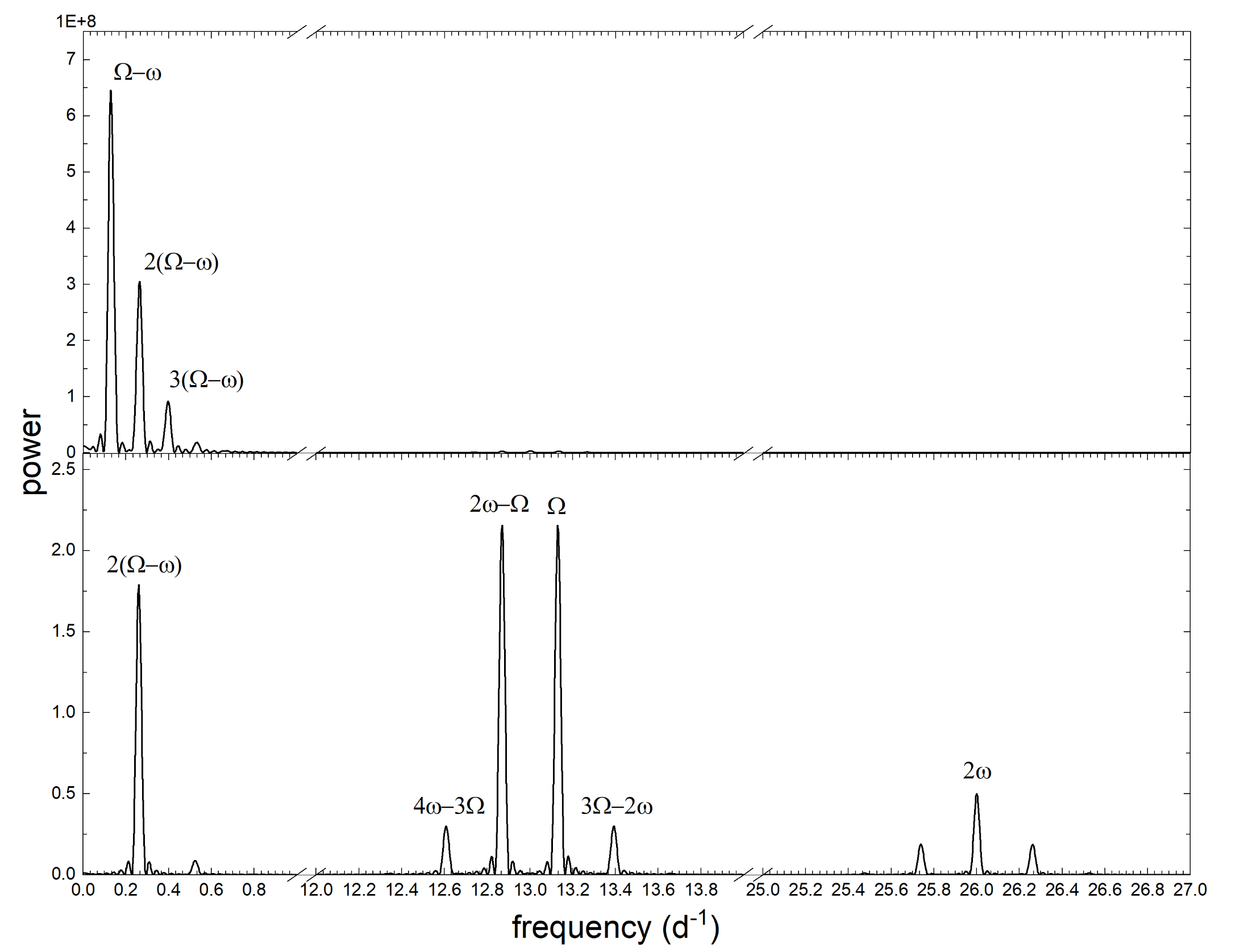}
\caption{Power spectra for dipole geometry with $\omega=13.00 d^{-1}$, $\Omega=13.13 d^{-1}$ and $m=20^{\circ}$, $\phi=0$ for the upper spot. The upper spectrum is the example of class 1 for $i=20^{\circ}$, and the lower spectrum, the example of class 2, for $i=90^{\circ}$. \label{fig_power_spectra_class1-2}}
\end{figure}

\begin{figure}[ht!]
\plotone{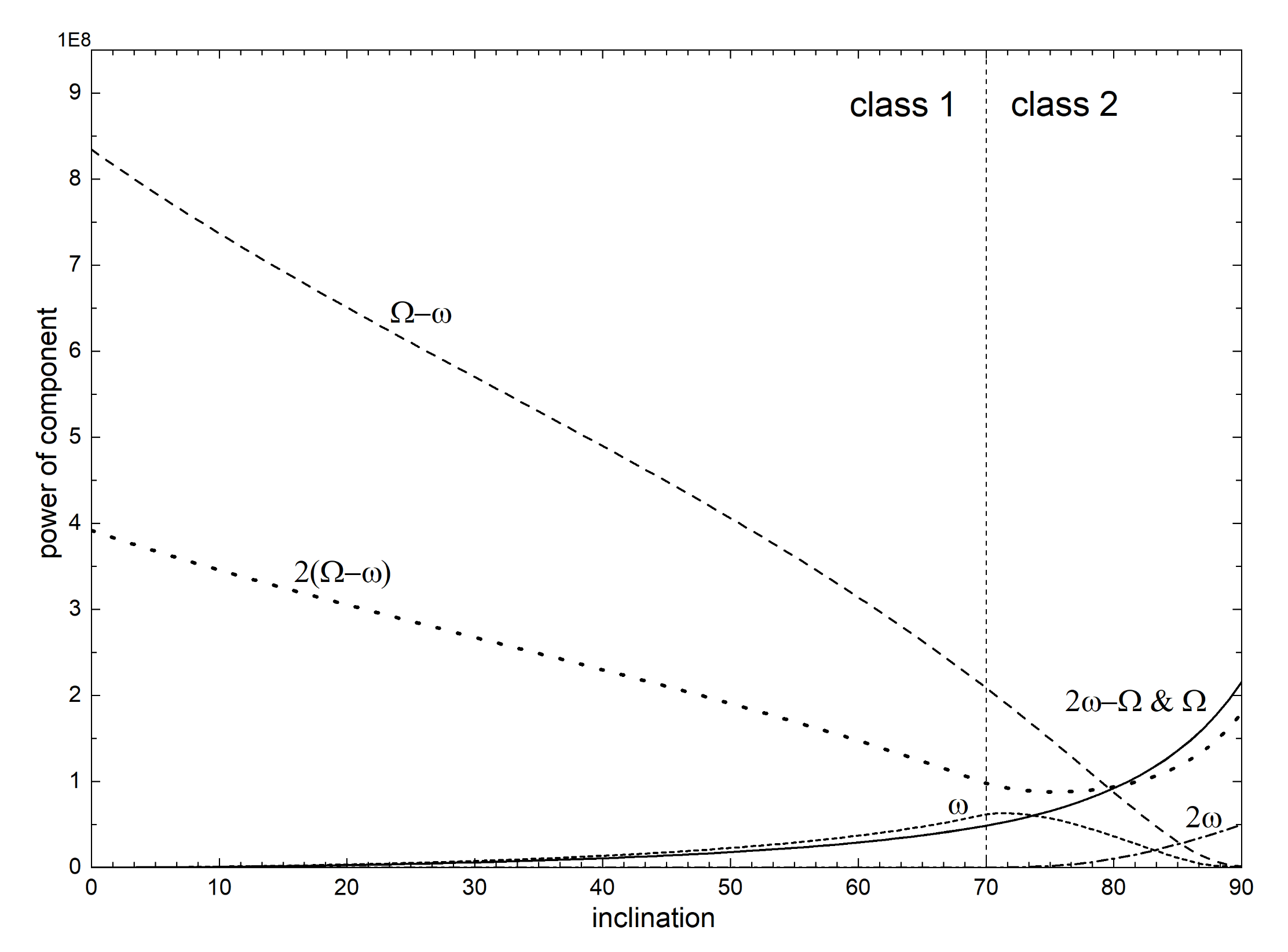}
\caption{The relation between the power of component and inclination for dipole geometry with $\omega=13.00 d^{-1}$, $\Omega=13.13 d^{-1}$ and $m=20^{\circ}$, $\phi=0$ for the upper spot. The vertical dash line at $70^{\circ}$ separates the two classes. Note that the power at $\Omega$ is same as that at $2\omega-\Omega$ for all inclinations. \label{fig_dipole_power_inclination}}
\end{figure}

Wynn \& King (1992) have studied the patterns of the power spectra for two classes of the dipole configuration with two spots diametrically opposed ($m$, $\phi$ for the upper spot and $\pi-m$, $\pi+\phi$ for the lower spot). Following their definitions, we presented the power spectra in Fig. \ref{fig_power_spectra_class1-2} for the examples of class 1 (one spot is completely visible while the other obscured at all phases) and class 2 (one spot alternates with the other one), and found some distinctive results given by the new model. A strong beat peak and its harmonics are visible in the class 1, while the class 2 spectrum shows two peaks with same power at orbital ($\Omega$) and side-band ($2\omega-\Omega$) frequency, only the second harmonics of the beat and spin frequency spike out. With the increasing of the inclination from $i=0^{\circ}$ to $i=90^{\circ}$ (see Fig. \ref{fig_dipole_power_inclination}), the powers at beat frequency and its second harmonic reduce shapely whereas that at the orbital frequency climb slowly, and the peak at spin frequency is only prominent at the transit region of the two classes. 

\begin{figure}[ht!]
\plotone{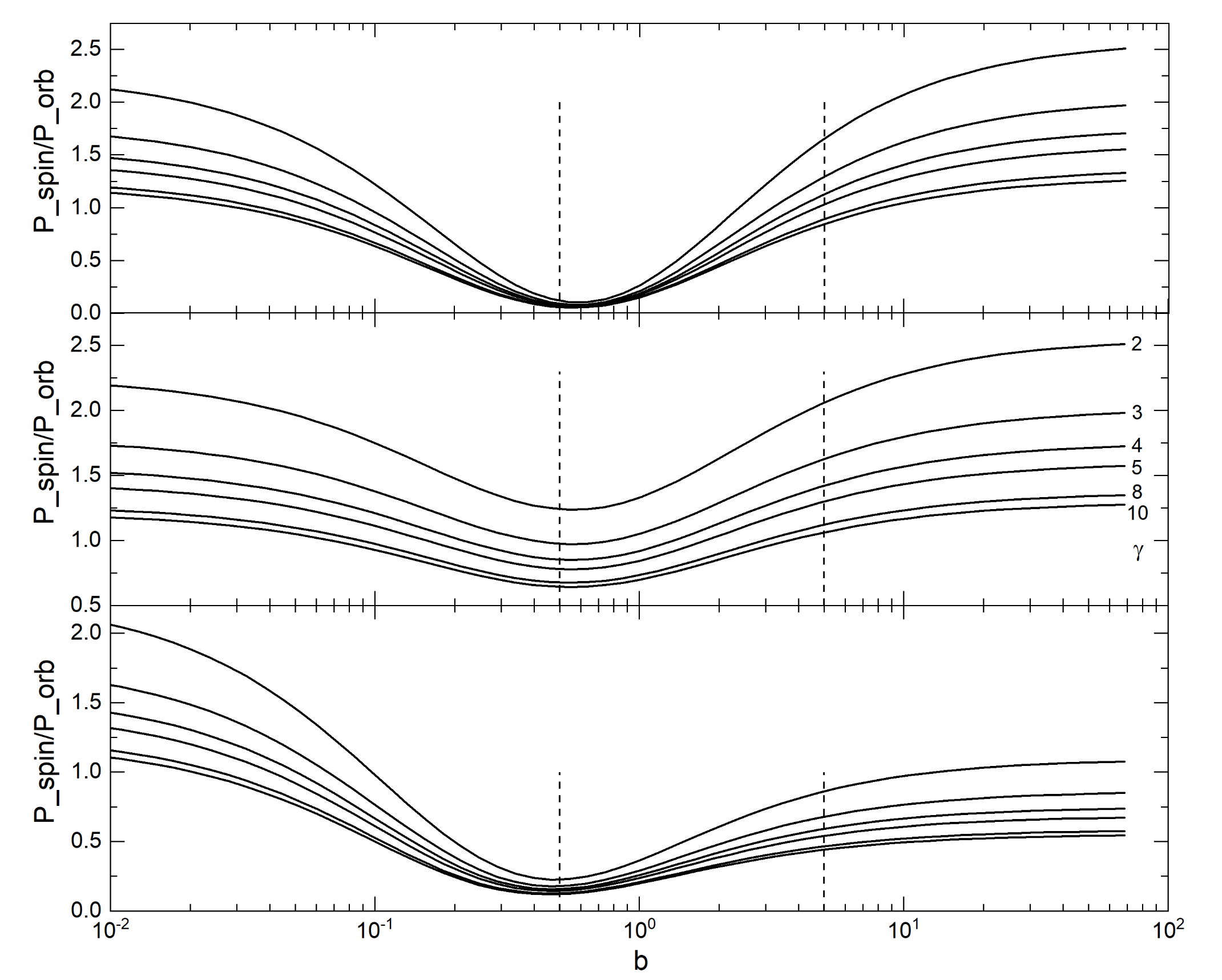}
\caption{The relations between $P_{spin}/P_{orb}$ and $b~(=B_{[2,3]}/B_1)$ for three kind of geometries using different $\gamma s$. The top panel is based on a geometry with two spots at $m_1$, $\phi_1$ and $m_3$, $\phi_3$ respectively; the middle one has two spots at $m_1$, $\phi_1$ and $m_2$, $\phi_2$ respectively; whereas the bottom one is the combination of these two geometries above, the $\gamma s'$ order in each panel is the same as the middle one. The values of all parameters are listed in Table \ref{tab:model}. Note that Spot 2 and 3 have same luminosity and co-latitude while their phase separation is 0.25. The vertical dash lines mark the $b=0.5, 2$. 
\label{fig_ReativeIntensityPspin2Porb}}
\end{figure}

However, we can not find a similar frequency distribution in the period analysis of light curves of APs like that presented by these two classes. The asynchronization of each AP is very slight ($(\Omega-\omega)/\Omega<2\%$, refer to \cite{Myers2017CDInd_Period}), so different frequencies cluster around the $\omega$, while the period analysis of the photometrical data often shows one peak spiking out there. As indicated in the Appendix \ref{appendixA:phase_separation}, the phase separation between two spots less than $\pi$ will make the power of orbital frequency larger than that of the side-band of $2\omega-\Omega$. So the relative power $P_{spin}/P_{orb}$ of the peaks at $\omega$ and $\Omega$ is very crucial for the identification of the WD spin period and the binary orbital period. The relations between the relative power $P_{spin}/P_{orb}$ and the relative visible brightness $b$ for different geometries are given in Fig. \ref{fig_ReativeIntensityPspin2Porb}; the top panel is based on a geometry with two spots at $m_1=30^{\circ}$, $\phi_1=0$ and $m_3=90^{\circ}$, $\phi_3=0.5$ respectively; the middle one has two spots at $m_1$, $\phi_1$ and $m_2=90^{\circ}$, $\phi_2=0.25$ respectively, whereas the bottom one is the combination of two geometries above. The other parameters of these geometries are listed in Table \ref{tab:model}. These relations impose an upper limit on the $P_{spin}/P_{orb}<2$ under the condition $\gamma>3$ and a more practical one is $P_{spin}/P_{orb}<1.7$ for two spot mode considering $b\sim1~(0.5..2)$. The combination of different geometries for any system will reduce the $P_{spin}/P_{orb}$. Note that, in order to reduce the dependence on the forms of the feeding intensity and changing aspect, the powers at different frequencies were calculated using the multi-term general Lomb-Scargle Method (GLSM, \cite{Lomb1976lomb-scargle, Scargle1982lomb-scargle, Zechmeister2009periodograms}) and here 2-term GLSM is enough for period analysis.

\begin{deluxetable*}{ccccccccccccc}
\tablenum{1}
\tablecaption{Parameters of the spot model}
\tablehead{$\Omega(d^{-1})$ & $\omega(d^{-1})$ & i & $phi_{L1}(2\pi)$ & $m_1$ & $\phi_1(2\pi)$ & $L_1$ & $m_2$ & $\phi_2(2\pi)$ & $L_2$ & $m_3$ & $\phi_3(2\pi)$ & $L_3$}
\startdata
13.13 & 13.00 & $70^{\circ}$ & 0.0 & $20^{\circ}$ & 0.0 & 1 & $90^{\circ}$ & 0.25 & L & $90^{\circ} $ & 0.5 & L
\enddata
\tablecomments{The $Ls$ present the relative luminosity and $\phi$s are in unit of $2\pi$.
\label{tab:model}}
\end{deluxetable*}

\section{CD Ind} \label{sec:cdind}
\begin{figure}[ht!]
\plotone{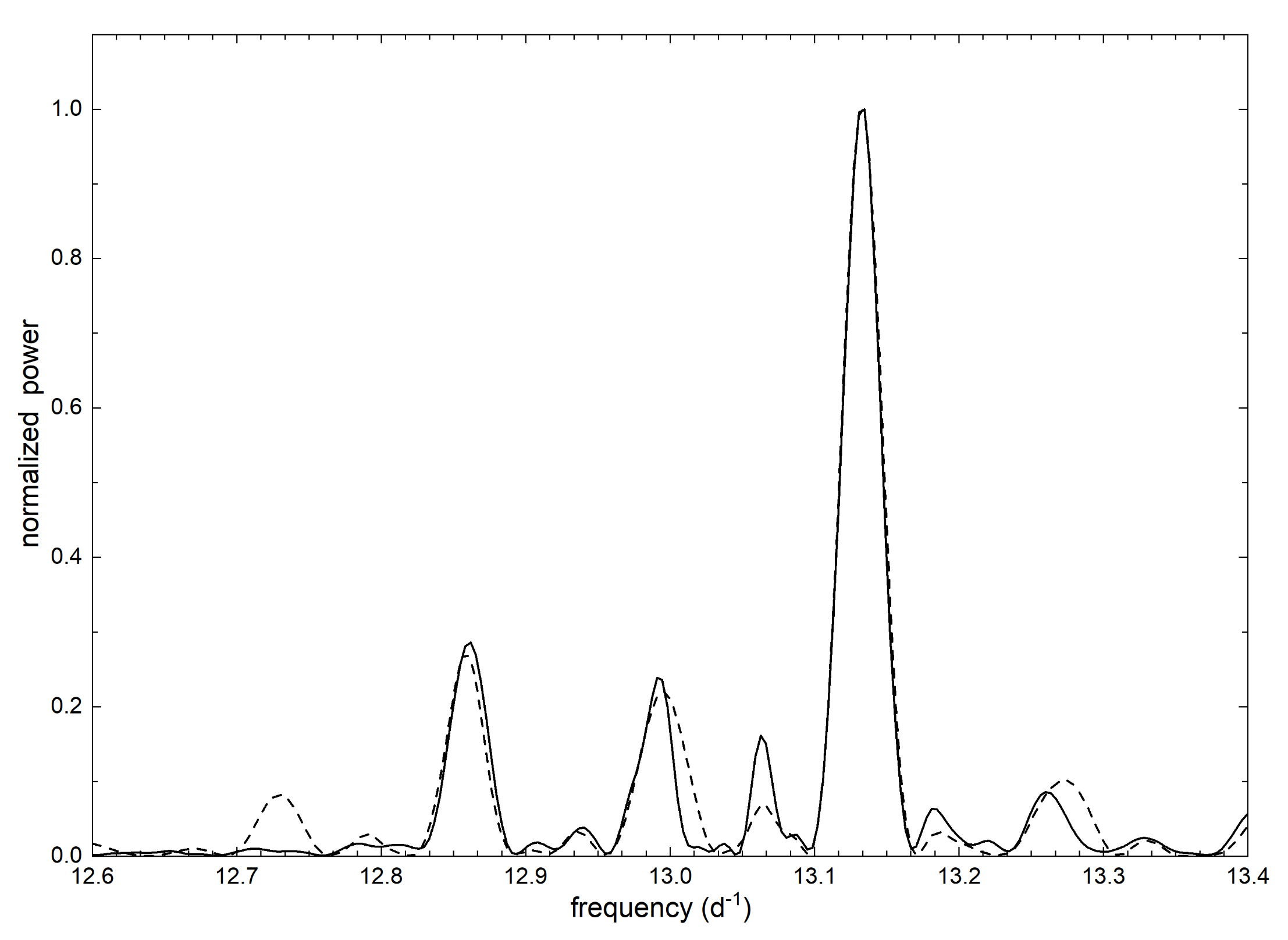}
\caption{The power spectra of TESS data and simulated optical data for CD Ind near the frequency $13 d^{-1}$. They are drawn by solid and dash lines respectively and are approximately coincident although there are some deviations, which implies that the accretion process is much unstable. The power spectra are normalized to the same level according to the strongest peaks. 
\label{fig_power_spectra_cdind}}
\end{figure}

\begin{figure}[ht!]
\plotone{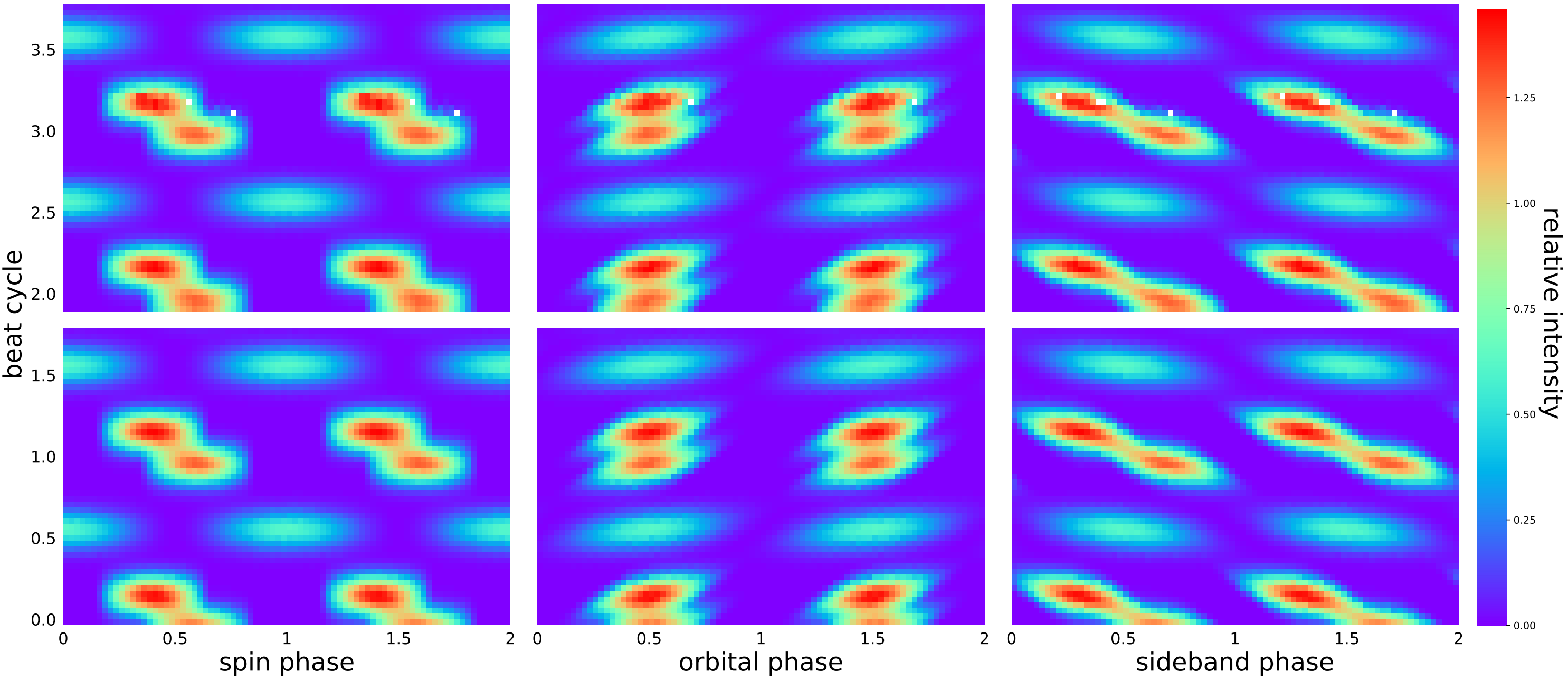}
\caption{Two-dimensional light curve of the simulated data for CD Ind, phased to the spin, orbital and side-band of $2\omega-\Omega$ frequencies, using a 0.5-d sliding window. The initial epoch is arbitrary for each phased data. Note that the humps (the hot spots) are much stabler along the beat cycle when phased to the orbital phase.
\label{fig_twoD_LC_CDInd}}
\end{figure}

\cite{Ramsay1999NewAp_CDInd} confirmed CD Ind as an AP from 2-week optical polarimetry and took $\omega=13.13 d^{-1}$ and $\Omega=13.00 d^{-1}$, then they found the polarization humps were at roughly similar spin phases. While \cite{Littlefield2019CDInd_TESS} identified the strongest frequency $13.13 d^{-1}$ as $2\omega-\Omega$ in the framework of Wynn \& King's model from the analysis of the optical data, which has been obtained through The Transiting Exoplanet Survey Satellite (TESS), and the features of its power spectra were displayed vividly by them, such as their Figure 2 and Figure 6, where we can pick up some parameters for the spot model. The lower accretion region was fed by the stream at the first half beat phase and then the upper region followed for the other half one. The emission from these two regions presents sinusoidal variation, albeit with the half cycle for the lower region trimmed off, which gives the support for one assumption in the model. Also the TESS light curve in the first eight days shows three humps during a beat cycle, which implies that there are three accretion regions and that the WD may have a complex magnetic field structure. So we adopt for CD Ind the geometry with $i=70^\circ$, one upper spot at $m=20^\circ, \phi=0$, and two lower spots at $m=90^\circ, \phi=0.4$ and $m=90^\circ, \phi=0.6$ respectively. The power spectrum of the TESS data near $13d^{-1}$ shows three peaks (see Fig. \ref{fig_power_spectra_cdind}). If the strongest peak is the WD spin signal the $P_{spin}/P_{orb}>3.4$ contradicts the prediction of the spot model. Also the model rules out the possibility of $P_{2\omega-\Omega}/P_{orb}>1$, so we regard the strong peak as the binary orbital signal. The Gaussian fitting to the peaks gives $\Omega=13.1327 d^{-1}$ and $2\omega-\Omega=12.8604 d^{-1}$, then $\omega=12.9965 d^{-1}$ is calculated from them rather then fitting the its peak due to the asymmetric shape. The other parameters of the geometry were listed in Table \ref{tab:cdind}. The power spectrum of the simulated data, sampled on the observation windows of the TESS data, near the frequency $\omega$ displays a similar distribution as the real one. The deviation between both spectra implies that the accretion process is very unstable. But we think it is enough to identify which one is the orbital period. Fig. \ref{fig_twoD_LC_CDInd} is the two-dimensional light curves of the simulated data, phased to the spin, orbital and side-band of $2\omega-\Omega$ frequencies, using a 0.5-d sliding window. Note that the humps are rather stable along the beat cycle when phased to the orbital phase. 

\begin{deluxetable*}{ccccccccccccc}
\tablenum{2}
\tablecaption{Parameters of the spot model for CD Ind}
\tablehead{
$\Omega(d^{-1})$ & $\omega(d^{-1})$ & i & $\phi_{L1}(2\pi)$ & $m_1$ & $\phi_1(2\pi)$ & $L_1$& $m_2$ & $\phi_2(2\pi)$ & $L_2$& $m_3$ & $\phi_3(2\pi)$ & $L_3$}
\startdata
13.1327 & 12.9965 & $70^{\circ}$ & 0.50 & $20^{\circ}$ & 0.0 & 1 & $90^{\circ}$ & 0.4 & 1.4 & $90^{\circ}$ & 0.6 & 1.6
\enddata
\tablecomments{Same as Table \ref{tab:model}. \label{tab:cdind}}
\end{deluxetable*}

\section{BY Cam} \label{sec:bycam}
\begin{figure}[ht!]
\plotone{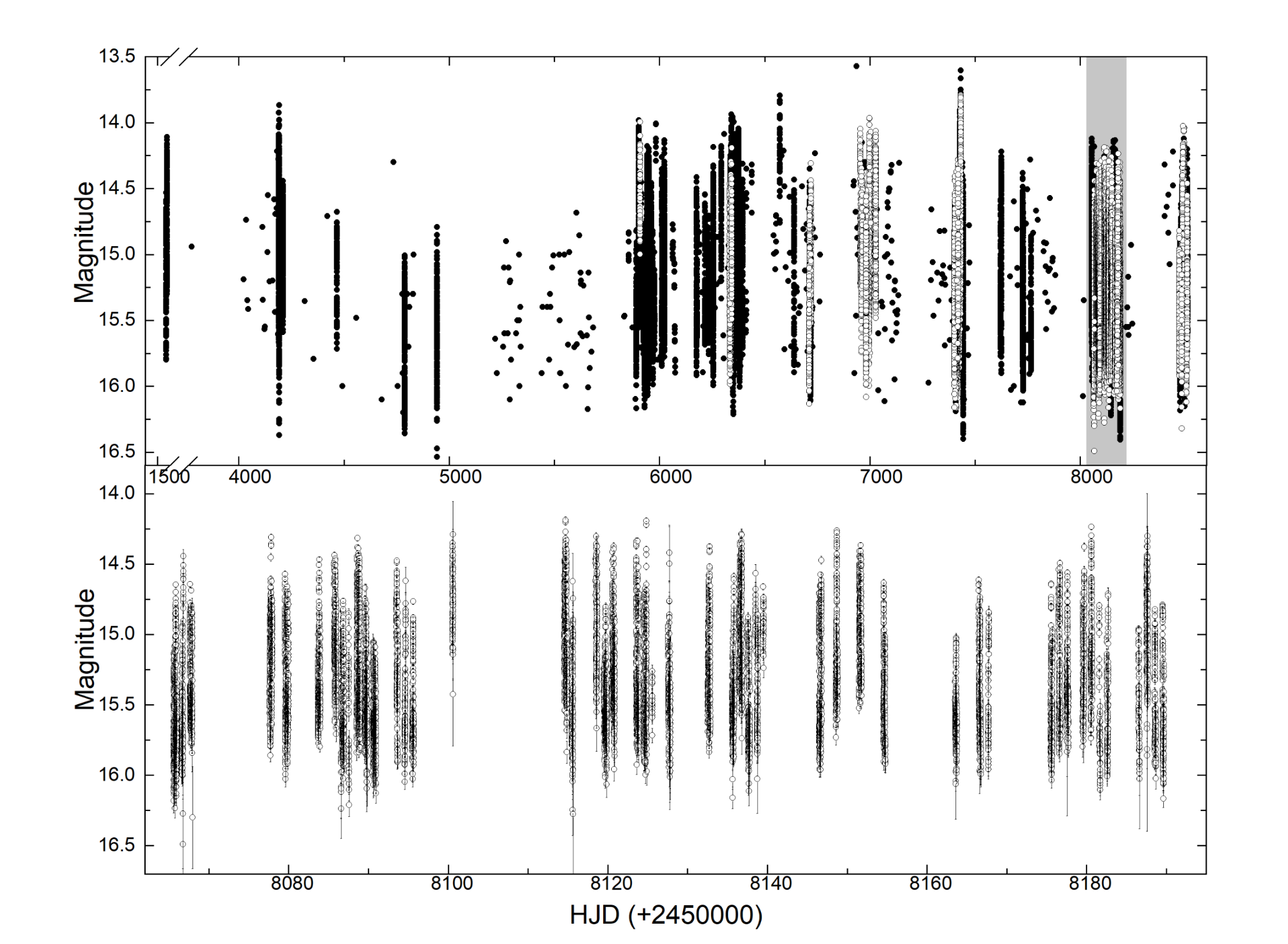}
\caption{The AAVSO data of BY Cam in CV band (the top panel). The open circles indicate the UJHA's data of which the part marked by the shadow column is presented in bottom panel; the solid circles represent the data obtained by others.
\label{fig_AAVSO_CV}}
\end{figure}

BY Cam has been observed by the American Association of Variable Star Observers (AAVSO)\footnote{https://www.aavso.org/} for almost 23.5 years since 1995 and those observations, in different bands, are contributed by different observers, for example, Ulowetz Joseph (UJHA) observed the star for 19848 times through CV (wide band V mag) band from 2011/12 to 2019/01. The top panel of Fig. \ref{fig_AAVSO_CV} shows all AAVSO data in CV band. The data with many scattered dots is separated by the gap of observing window and shows the intensity of system changed during the observation seasons. Considering the stability and uniformity, only the part of UJHA data with HJD ranging from 2458060 to 2458200 was used for period analysis. This part is denoted as CV\_UJHA data and depicted in the bottom panel. The CV\_UJHA data is mostly built up from a series of sections of which most are up to an orbital period and have the temporal resolution of about 1.4 minutes. The data errors are typically 0.03 mag, and also presented in the bottom panel.
 
\begin{figure}[ht!]
\plotone{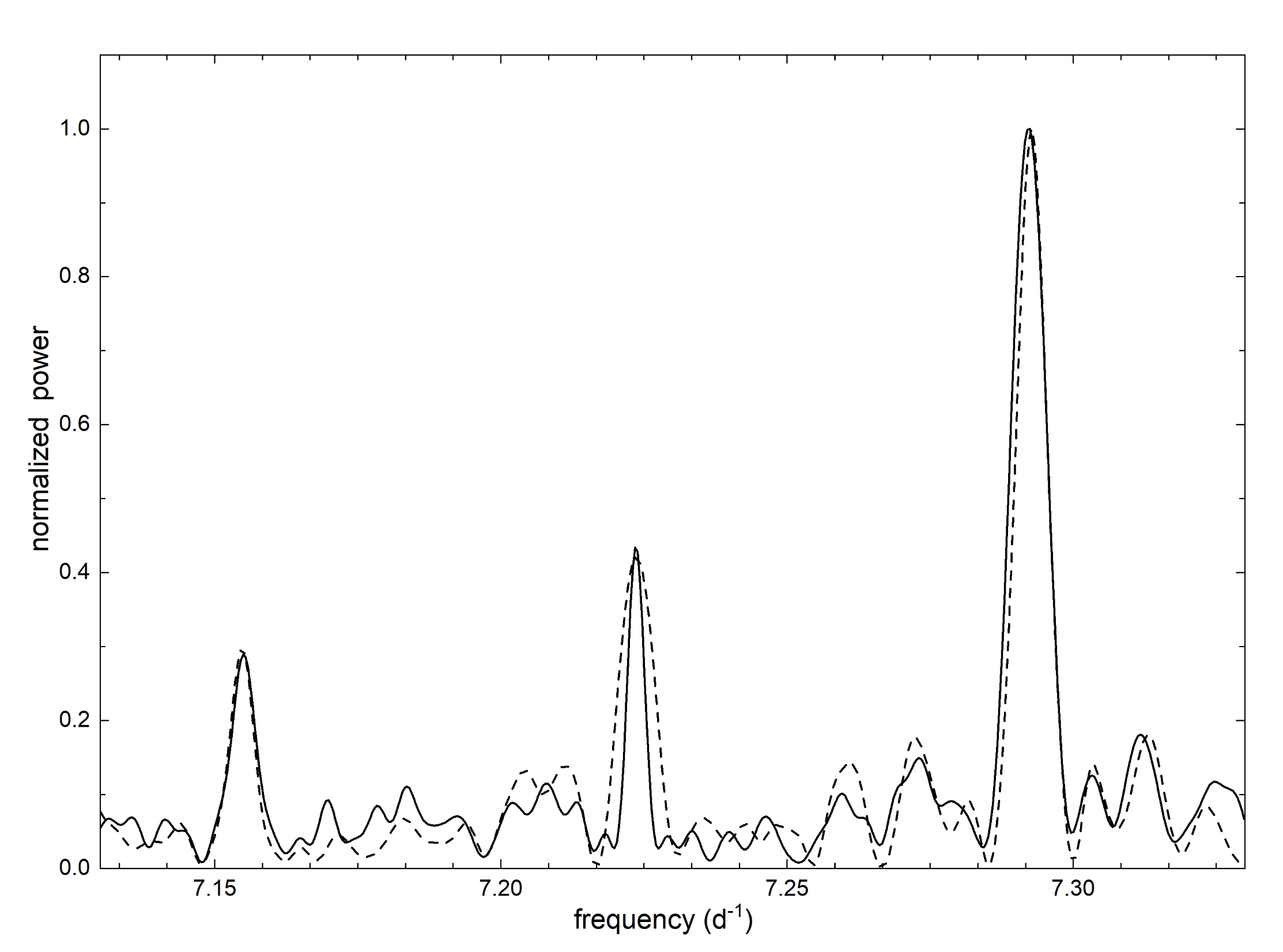}
\caption{The power spectra of CV\_UJHA data (the solid line) and simulated optical data (the dash line) for BY Cam near the frequency $7.22 d^{-1}$.  The power spectra are normalized to the same level according to the strongest peaks. 
\label{fig_power_spectra_bycam}}
\end{figure}

In general, there are three frequencies $7.15 d^{-1}$ (=f3), $7.22 d^{-1}$ (=f2) and $7.29 d^{-1}$ (=f1) near the interesting frequency region, those are assigned as the orbital frequency, the white dwarf spin and the side-band of $2\omega-\Omega$, respectively. However, \cite{Honeycutt2005BYCam_orbperiod} assigned the f1 to orbital frequency because of its stability. At this point, the assignments of those frequencies is still uncertain. The spectrum of CV\_UJHA data depicted in Fig. \ref{fig_power_spectra_bycam} gives $P_{f1}/P_{f2}>2.3$, which supports that the f1 is the real orbital frequency. The simulation result was also shown in Fig. \ref{fig_power_spectra_bycam} using the geometry with the parameters given in Table \ref{tab:bycam}.

\begin{deluxetable*}{ccccccccccccc}
\tablenum{3}
\tablecaption{Parameters of the spot model for BY Cam}
\tablehead{$\Omega(d^{-1})$ & $\omega(d^{-1})$ &i  & $\phi_{L1}(2\pi)$ & $m_1$ &$\phi_1(2\pi)$ & $L_1$ & $m_2$ & $\phi_2(2\pi)$ & $L_2$ & $m_3$ & $\phi_3(2\pi)$ & $L_3$}
\startdata
7.2926 & 7.2236 & $70^{\circ}$ & 0.6 & $50^{\circ}$ & 0.0 & 1 & $140^{\circ}$ & 0.5 & 1 & $90^{\circ}$ & 0.78 & 0.6
\enddata
\tablecomments{Same as Table \ref{tab:model}.
\label{tab:bycam}}
\end{deluxetable*}

\section{1RXS J083842.1-282723} \label{sec:RXJ0838-2827}
\begin{figure}[ht!]
\plotone{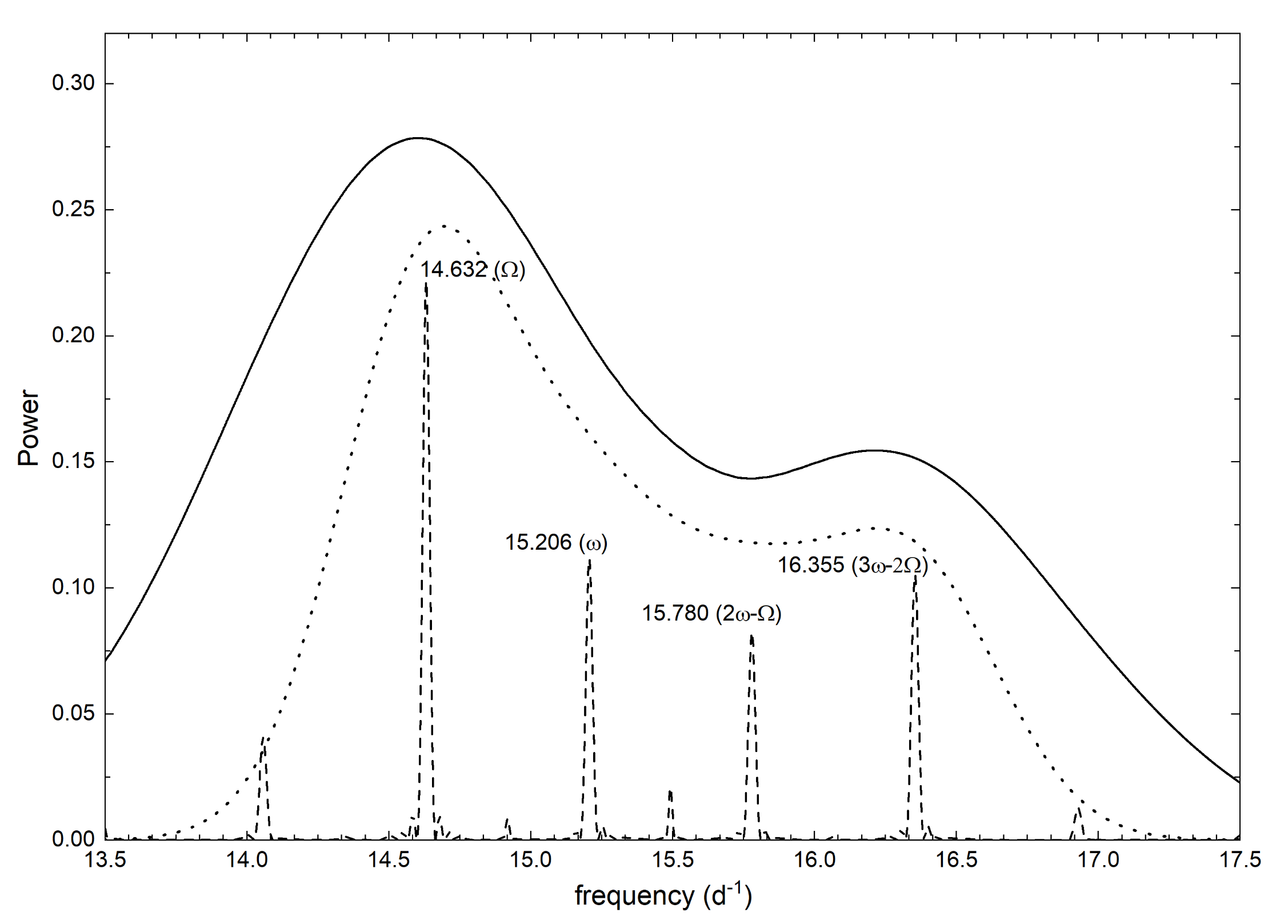}
\caption{The power spectrum (dashed line) of the simulation data for RX J0838 shows four strong peaks marked by the labels. A convolution (dotted line) of these peaks with a gaussian kernel of $\sigma=0.3d^{-1}$ is over-plotted on the spectrum of the XMM-Newton data (solid line). The comparison of them shows that the 1.47h-signal corresponding to 16.355 $d^{-1}$ peak is the side-band of $3\omega-2\Omega$.
\label{fig_power_spectra_RXJ0838}}
\end{figure}

1RXS J083842.1-282723 (RX J0838 hereafter) shows several Balmer and helium emission-lines in optical spectrum and was tentatively classified as a CV by \cite{Masetti2013ROSATFermi}. Based the optical and X-ray observations, \cite{Halpern2017J0838-2829} found RX J0838 is a strongly AP in which accretion switches between two poles that are $120^{\circ}$ apart and that the X-ray light curve modulates at 94.8(4) minutes and the optical emission line varies at period of 98.413(4) minutes, these two periods were interpreted as the white dwarf spin period and the orbital period, respectively. While the power spectrum of the X-ray data has a peak at 1.47 hours which was identified by \cite{Rea2017J0838-2829} as the spin period. In order to find where this signal comes from, we simulated the light curve of RX J0838 with a simply dipole accretion geometry and, considering the large gap in the X-ray data, adopted a dense observation windows. The power spectrum (see Fig. \ref{fig_power_spectra_RXJ0838}) with the parameters of the geometry listed in Table \ref{tab:RXJ0838} shows 4 strong peaks. A convolution of these 4 peaks with a Gaussian ($\sigma=0.3d^{-1}$) is calculated to add the effect of the big window gap, which shows that the side-band of $3\omega-2\Omega$ at 16.355 $d^{-1}$ causes the 1.47h-signal shown up in the power spectrum of the X-ray data.  

\begin{deluxetable*}{cccccccccc}
\tablenum{4}
\tablecaption{Parameters of the spot model for RX J0838}
\tablehead{
$\Omega(d^{-1})$ & $\omega(d^{-1})$ & i   & $\phi_{L1}(2\pi)$ & $m_1$ & $\phi_1(2\pi)$ & $L_1$ & $m_2$ & $\phi_2(2\pi)$ & $L_2$}
\startdata
14.632 & 15.206 & $70^{\circ}$ & 0.2 & $30^{\circ}$ & 0.0 & 1 & $150^{\circ}$ & 0.333 & 3
\enddata
\tablecomments{Same as Table \ref{tab:model}.
\label{tab:RXJ0838}}
\end{deluxetable*}

\section{Discussion and Summary} \label{sec:Summary}
The spot model, plus the using of the multi-term GLSM, can be applied for the period analysis of the optical, X-ray or even other band light curves, only if there exist the modulations due to the changing aspect and the variable feeding intensity. The forms of these effects and the shapes of the accretion regions have little effects on the frequency distribution of the light curves. Indeed the light curves of APs are much more erratic, which indicates that the pole switching is way unstable and that the accretion process even changes during the course of the beat epoch. Considering the lack of knowledge of the accretion process between the mass and the magnetic field, a more physical model is unrealistic. Although the spot model is a phenomenological model, it could provide a glimpse of the accretion process. The light curves of CD Ind and RX J0838 show that the feeding intensity of the active spot is dependent on the position of the L1 point with respect to the spot. The light curves of BY Cam and CD Ind are much more variable than that of the RX J0838 and their accretion geometries are far deviation from the dipole configuration, which imply that the magnetic field is much more complex than that of synchronous polars. While RX J0838 is special not only for its simple accretion geometry and rather stable light curve but also for its highest asynchronism (see Table \ref{tab:aps}), which may indicate that this object is a pre-polar approaching its synchronization of the WD spin with the orbiting of its companion. Because the smooth valley of its light curve out the pole-switch phase, the times of the minimum are listed in Table \ref{tab:timesofminimunofRXJ0838}. More observations on RX J0838 will be valuable for the study of the evolution of the pre-polar and helpful for modeling the accretion process between the field and the matter due to its simple accretion geometry.

Based on the spot model, we identified the periods of APs. These periods plus that of other APs are listed in Table \ref{tab:aps} where V1500 Cyg and RX J3808 are over-synchronous polars. While maybe RX J3808 is a pre-pole rather than an APs. And the spot model gives a strong criterion about the relative power $P_{spin}/P_{orb}<2$, which is very helpful for the identification of the periods. Considering the variation of accretion process, the power spectrum of long-term light curve will make the orbital period prominent.

\begin{deluxetable*}{cccc}
\tablenum{4}
\tablecaption{The Asynchronous Polars}
\tablehead{Systems & $P_{orb} (h)$ & $P_{spin} (h)$ & $(P_{orb}-P_{spin})/P_{orb}$}
\startdata
V1500 Cyg&3.3507[1]  &3.2917[1]& 0.0176\\
V1432 Aql&3.3657[1]  &3.3751[1]&-0.0029\\
CD Ind   &1.8275[2,3]&1.8470[3]&-0.0106\\
BY Cam   &3.2910[4,5]&3.3224[5]&-0.0095\\
RX J0838 &1.6402[6]  &1.58  [6]& 0.037\\
\enddata
\tablecomments{
[1] \cite{Ramsay1999NewAp_CDInd};
[2] \cite{Myers2017CDInd_Period};
[3] The TESS data;
[4] \cite{Honeycutt2005BYCam_orbperiod};
[5] The CV\_UJHA data;
[6] \cite{Halpern2017J0838-2829}
\label{tab:aps}}
\end{deluxetable*}

\begin{deluxetable*}{cccc}
\tablenum{5}
\tablecaption{The BJDs of the minimum in the light curve of the RX J0838}
\tablehead{BJDs & Errors & Cycles & Types[a]}
\startdata
2457316.00304&0.00008&1&I\\
2457316.06852&0.00006&2&I\\
2457316.13467&0.00006&3&I\\
2457316.20114&0.00008&4&I\\
2457359.61024&0.00006&1&II\\
2457359.67573&0.00004&2&II\\
2457359.74224&0.00002&3&II\\
2457359.80878&0.00004&4&II\\
2457359.87562&0.00007&5&II\\
\enddata
\tablecomments{
[a] I means a certain accretion spot is active while II means the other spot is active.
\label{tab:timesofminimunofRXJ0838}}
\end{deluxetable*}

\acknowledgments
This work is partly supported by Chinese Natural Science Foundation (No. 11933008 and No. 11803083). This paper includes data collected by the TESS mission, which are publicly available from the Mikulski Archive for Space Telescopes (MAST). Funding for the TESS mission is provided by NASA’s Science Mission directorate. We acknowledge with thanks the variable star observations from the AAVSO International Database contributed by observers worldwide and used in  this  research. The work is also based on observations obtained with XMM–Newton, an ESA science mission with instruments and contributions directly funded by ESA Member States and the USA (NASA).

\appendix
\section{The effect from the phase separation less than pi} \label{appendixA:phase_separation}
In the appendix, we will present the effect of the two emission regions non-diametrically opposed. As mentioned in the text, there are two effects that can modulate the emission from an accretion spot, then the radiation from the spot A at time $t$ can be simulated simply as
\begin{equation}\label{A1:spotA_signal}
l_A(t)=L_A\cos{(2\pi\mu t+\phi_A)}\cos{(2\pi\nu t+\phi_{AL})},
\end{equation}
where $\mu=\omega$, $\nu=\omega-\Omega$ and $\phi_{AL}=\phi_{A}-\phi_{L}$. The Fourier transform of Eq.\ref{A1:spotA_signal} gives the distribution of the frequencies
\begin{equation}\label{A2:fourier_A1}
\mathscr{F}[l_A(t)]=
\frac{L_A}{4}\left(
\delta_{\mu+\nu}e^{i(\phi_A+\phi_{AL})}+
\delta_{\mu-\nu}e^{i(\phi_A-\phi_{AL})}
\right)+...
\end{equation}
where $\delta$ is Dirac function with $\delta_{\mu+\nu}\equiv\delta(t-\mu-\nu)$ and the terms relative to other frequencies are omitted. Eq.\ref{A2:fourier_A1} exhibits that the intensities of two frequencies at $\mu+\nu \ (2\omega-\Omega)$ and $\mu-\nu\ (\Omega)$ have same strength, just as the Wynn $\&$ King model said, which is disapprove of the face revealed in the power spectra. However, this conflict can be resolve by adding another emission region Spot B at $\phi_B$, of which the light curve is similar to that of Spot A:
\begin{equation}\label{A3:spotB_signal}
l_B(t)=L_B\cos{(2\pi\mu t+\phi_B)}\cos{(2\pi\nu t+\phi_{BL})}.
\end{equation}
And, after substituting $\phi_{AL}$ and $\phi_{BL}$, the Fourier transform for the new signal set, Eq.\ref{A1:spotA_signal}+Eq.\ref{A3:spotB_signal}, is
\begin{equation}\label{A4:fourier_A1+A3}
\begin{split}
\mathscr{F}[l_A(t)+l_B(t)]=
\frac{1}{4}\delta_{\mu+\nu}e^{i\phi_A}\left(L_A+L_Be^{2i\phi_{BA}}\right)+
\frac{1}{4}\delta_{\mu-\nu}e^{i\phi_L}\left(L_A+L_B\right)+...\\
\end{split}
\end{equation}
where $\phi_{BA}=\phi_{B}-\phi_{A}$. So the powers $P_{\mu+\nu}$ and $P_{\mu-\nu}$ at frequencies $\mu+\nu$ and $\mu-\nu$ are $(L_A/4)^2+(L_B/4)^2+2(L_A/4)(L_B/4)\cos{2\phi_{BA}}$ and $(L_A/4)^2+(L_B/4)^2+2(L_A/4)(L_B/4)$ respectively. When $\phi_{BA} = \pi$, $P_{\mu+\nu} = P_{\mu-\nu}$, which corresponds to the geometry of two diametrically opposed emission regions, while $\phi_{BA} < \pi$, $I_{\mu+\nu} < I_{\mu-\nu}$, which describes the effect of two regions separated in phase $<\pi (180^{\circ})$ and is the case revealed in the power spectra.

\bibliography{sample63}{}
\bibliographystyle{aasjournal}

\end{document}